# Agile data management in NAV: A Case Study

Kathrine Vestues[1], Geir Kjetil Hanssen[1], Marius Mikalsen[1,2], Thor Aleksander Buan[1], Kieran Conboy[1,3]

[1] SINTEF, Trondheim, Norway
`{kathrine.vestues,geir.k.hanssen,marius.mikalsen,thor.buan}`
`@sintef.no`
[2] Norwegian University of Science and Technology, Trondheim, Norway
`marius.mikalsen@ntnu.no`
[3] Lero Centre for Software Research, NUI Galway, Ireland
`kieran.conboy@nuigalway.ie`

**Abstract.** To satisfy the need for analytical data in the development of digital services, many organizations use data warehouse, and, more recently, data lake architectures. These architectures have traditionally been accompanied by centralized organizational models, where a single team or department has been responsible for gathering, transforming, and giving access to analytical data. However, such centralized models presuppose stability and are incompatible with agile software development where applications and databases are continuously updated. To achieve more agile forms of data management, some organizations have therefore begun to experiment with distributed data management models such as "data meshes". Research on this topic is however limited. In this paper, we report findings from a case study of a public sector organization in Norway that has begun the transition from centralized to distributed data management, outlining both the benefits and challenges of a distributed approach.

**Keywords:** Agile software development, distributed data management, Data mesh, empirical, case study

## 1 Introduction

Most software organizations are aiming to become data driven, where all business units take an active role in both the production and consumption of analytical data. However, this "democratization" of data [1] challenges traditional centralized data management architectures and organization models, such as the data warehouse [2]. Data warehouse models, where a single team or department is responsible for managing analytic data require predictability and stability, characteristics which are incompatible with agile development.

A common challenge within data management is that the logic and flow of data does not follow the structure of the organization [3]. For instance, centralized data management does not follow the logic of agile software development where autonomous, cross-functional teams have end-to-end responsibility for products. Mismatches between organizational structures and data usage can lead to issues such as data silos and unclear responsibilities. This is especially problematic when developing analytical solutions that cross organizational boundaries and rely on data from different silos [3]. A proposed remedy gives teams increased ownership of data produced by their applications [4]. Such initiatives aim to improve the coordination of people, process, and technology to enable more agile and automated approaches to data analytics [4] [5]. The goal is to bring stakeholders such as data architects, data engineers, data scientists, application developers and data consumers together in building analytical solutions in an agile and collaborative manner [6].

One approach to agile data management is "data-mesh" [7]. Dehghani [7], [8] describes the data mesh in terms of four core principles: 1) domain-oriented decentralized data ownership and architecture, 2) data as a product, 3) self-serve data platform, and 4) federated computational governance. Unlike central management models, such as data warehouse or data lake, the data mesh sees data as context-dependent and best managed in a distributed manner [9]: Those who produce and know the data are best equipped for curating and distributing it.

While there is a rich body of literature on data management, focusing on areas such as the collection, curation, consumption and control of data, empirical papers describing distributed data management and data mesh are still scarce. This is problematic, considering the emphasis among researchers and practitioners on increasing the use of analytical data in improving the efficiency and quality of services. We therefore ask the following research question: *What are the challenges for agile software development organizations when introducing distributed data management?*

We seek to answer this question by reporting findings from an interpretive case study of the development unit in NAV, short for the Norwegian Labor and Welfare administration. NAV forms the backbone of the Norwegian welfare system and is responsible for redistributing one third of the national budget through schemes such as age pension, sick benefit, and unemployment benefit. To provide analytical insight both inside and outside of NAV, data has been collected and curated by a centralized unit and processed in a data warehouse consisting of many registers. Whereas the centralized model worked satisfactory in a system landscape with large systems that rarely changed, it has proven problematic as the organization transitions towards agile development teams and continuous development. To address these challenges, NAV has begun the implementation of a distributed data management model, inspired by the principles of data mesh [7].

Our study sheds light both on the potential benefits of distributed data management, as well as the challenges that such approaches cause. The findings are a first step towards a process model capturing the transition from centralized to decentralized data management. It will also assist practitioners who consider a similar change.

The rest of the article is organized as follows: Section 2 presents the case background and the methods, while section 3 presents the findings. Section 4 discusses the findings and outlines some key challenges that must be solved. Section 5 concludes with a consideration of future possibilities for research.

## 2 Background

### 2.1 Analytics

In order to study analytics the first step is to provide a definition of what the term means in the context of this study. Analytics are frequently referred to as the techniques, technologies, systems, practices, methodologies, and applications that enable organisations to analyse critical business data [10]. Seddon and Currie [11] propose a definition that is concerned with evidence-based problem recognition and solving that occur within the context of business environments, namely *"the use of data to make sounder, more evidence-based business decisions"*. This is the definition adopted in this study. However, the extant conceptualisation and classification of business analytics is quite limited and what does exist [12][13][11] tends to vary greatly. In terms of getting to a more specific and operationalised definition of business analytics that can be used, this study draws on [13], which systematically reviewed and consolidated the extant conceptualisations of business analytics. Their literature review showed that, in terms of describing the data characteristics that underpin the notion of business analytics, many characteristics exist; however, the three key attributes include *volume*, *velocity* and *variety* of data [14][15]. Given that this is an exploratory study, we chose to adopt a broader perspective regarding the data attributes that are relevant in business analytics.

### 2.2 Analytics in the public sector

The Norwegian public sector is highly digitalized and represents a data-rich domain with access to advancing technologies for analyzing and utilizing data. This underpins the idea of a "data-driven" public sector where data analytics are seen as a path to better policymaking and improved services [16]. However, business analytics can also be challenging. In a study of the Norwegian public sector, Broomfield and Reutter [16] identified several challenges. Among these were: 1) *Organizational culture*, 2) *Privacy and security concerns,* 3*) Outdated legal and regulatory frameworks*, 4) *Data quality* - where the use for data analytical purposes may put additional requirements relating to contextualization, biases, and the suitability of data, and 5) *Access to data* - where data needs to be accessible, both from a technical and an organizational standpoint.

Although analytics in the public sector has become an established research field [17], especially from the organizational and regulatory perspective, the technical and IT perspectives are in a nascent state with few empirical studies available [18][19][20].

### 2.3 Agile analytics and data mesh

There are many debates emerging regarding the use of analytics in high speed, agile environments, e.g., the use of analytics in a democratized manner [1] or the use of analytics to enable dynamic capabilities [21]. From a practitioner literature, the concept of a "data mesh" [7], [8] has been proposed as a novel means of managing analytical data. Inspired by Eric Evans book on domain-driven design [22], Dehghani [7] argues that data should be built and managed around "domains", proposing 4 principles which will enable organizations to manage analytical data at scale: *1) Domain-oriented decentralized data ownership and architecture*: Data are owned, managed, and located according to their business or thematic domain, e.g., being the responsibility of domain teams that have deep insight in their domain, and then also the domain-oriented data. *2) Data as a product*: In much the same way as teams sees the software they produce as a product (typically in the form of a service) where they have a special responsibility to the end-users, data are also treated as a product. A data product must have the right level of quality and availability, where the owner understands the needs of the consumer of the data product. In practical terms, a data product consists of code (data pipelines to access data), data and metadata (the actual data and metadata that is needed to understand and use data), and infrastructure (to execute the code and to store the data). *3) Self-serve data platform:* In similar ways as teams may use a shared application platform to deliver their software products to consumers, they also need a platform to deliver their data products to consumers, such as other teams or data analysts. The platform offers tools and infrastructure for simplified provisioning as a shared asset in the organization. This can be infrastructure and tools for creating, maintaining, announcing, and sharing data products. *4) Federated computational governance:* Following the distribution of data and the responsibility for data comes a need for a federated approach to govern and improve the data mesh, including common principles and a shared data platform. Governance is a shared responsibility between data product owners, their consumers, and data platform product owners.

However, despite increasing attention among both researchers and partitions, there are to date few peer-reviewed empirical studies that exploring how agile data management and data mesh is addressed by organizations. Apart from informative whitepapers and internal presentations, e.g., from Zalando and Netflix [20], we have only identified one empirical study [18]. The reported transition indicates that the data mesh might increase analytical capabilities, suggesting that more industrial studies of practice are needed.

## 3  Research site and methods

### 3.1  Case background

The focus of the study is on how changes to organization, technology architecture, and software development approach is affecting the management of analytical data. The research was performed within the IT department of NAV, short for the Norwegian Labor and Welfare Administration. The IT department has approximately 800 employees that maintain and operate welfare services. The organization uses consultants as needed in development initiatives. NAVs IT system portfolio is made up of several generations of solutions, from mainframe systems to modern web-oriented applications, as well as standard systems that support operations such as accounting, payroll, and document production.

4## 3.2 Data collection

Data was collected from two main sources: Interviews and document reviews. To capture several aspects of the shift from centralized to decentralized data management, we chose informants from three parts of the organization: 1) data warehouse teams, 2) application development teams, and 3) the data platform team (the team responsible for developing the new data platform). Since these teams were cross-functional, they had members belonging both to the IT department (technical expertise) and to relevant business areas. These three categories of informants were chosen because they cover the various roles involved in data analytics within NAV: The data warehouse teams consumed data, the Application development teams produced data, while the data platform team developed the platform and facilitated the exchange of data. The number of informants, and their distribution across the different types of teams is listed in Table **1**.

Although the long-term goal is for application development teams to both produce and consume analytical data, this has not yet occurred. In this first stage of the transition, the organization's focus has been on supporting existing uses of analytical data, rather than using data in new ways.

**Table 1.** Overview of interviews.

|  |  | Team type | | |
| --- | --- | --- | --- | --- |
|  |  | 1) Data warehouse | 2) Application | 3) Data platform |
| Role | Data scientists |  | 2 |  |
|  | Data analysts | 1 |  |  |
|  | Developers | 3 | 3 |  |
|  | Privacy coach |  | 1 |  |
|  | Product owners | 1 | 2 | 3 |
|  | Managers |  |  | 2 |
|  |  | 5 | 8 | 5 |

We performed 18 semi-structured interviews. Of these, 12 were recorded and transcribed. Interviews lasted between 30 and 60 minutes. In the cases where we were unable to record the interviews, one researcher asked questions, while another took extensive notes. Informants were recruited through a snowballing approach, where one informant would suggest another. Typically, we were guided towards respondents that were known to have updated knowledge, competency, and interest in the topics that are relevant to our study, e.g., on the construction of the data platform, domain teams that are early adopters, data scientists looking for data, managers of groups that are impacted by the data mesh initiative, etc.

A second important source of data were document reviews. These documents included project steering documents, descriptions of the new data strategy (as proposed by the NAV IT department, online documentation of the data platform (GitHub), descriptions of NAV's IT ambition, and conference presentations held by members of the development organization (i.e., the Norwegian JavaZone conference[1], and the data mesh podcast[2]). Many of these presentations have been recorded and published online. They provided insight into the public version of NAV's IT and data strategies.

In addition to the data sources mentioned above, one of the authors studied the transformation of NAV's IT department from 2017 to 2019 as part of her PhD work [23]. In this period, NAV transformed its software development strategy and application architecture. Informants described this transformation as a trigger for the transition from centralized to decentralized data management. Changes to the software development strategy therefore needs to be seen in connection.

### 3.3 Data analysis

The data analysis can be described as an iterative three-step process [24]. In the first step, we explored appropriate literature to conceptualize the phenomenon of interest. Initially, we focused on the literature on open data. However, as we began the fieldwork, we learned that NAV's focus was on improved data sharing *inside* NAV. The rationale behind this internal focus was that effective data sharing with external partners, requires efficient data sharing internally. Attention was therefore shifted from external to internal data management, where we paid attention to the data mesh concept [7], which very clearly motivated the IT-organization

In the second step of data analysis, data was examined inductively through a manual coding process. Among the codes to emerge were "data product", "data platform", and "ownership". The codes were discussed and grouped into meaningful categories. We derived at two overarching categories, namely Centralized data management and Agile data management. We applied a manual approach for coding, where paper prints of transcripts and notes were shared between three of the researchers, sections that were found to exemplify or explain the implementation and viewpoints on the data mesh principles were extracted (cutting out text snippets) and arranged in groups that were given descriptive titles (codes).

In the third step, the inductively derived codes were merged with concepts from the literature. We found that our codes largely overlapping with the principles of "data mesh" [7], [8], leaving us with 3 categories of Agile data management: data ownership and products, data platform, and data governance. This provided us with structured insight into the organization's interpretation and adaption of data mesh.

---

[1] https://javazone.no
[2] https://daappod.com/data-mesh-radio/early-platform-insights-goran-berntsen-and-audun-fauchald-strand/

## 4 Findings

### 4.1 Background to the transition

To increase the efficiency and flexibility of public services, NAV has made substantial changes to the way they develop and disseminate software during the past few years. Handovers between departments have been replaced by continuous development, and hierarchical organization has been replaced by cross-functional teams that take responsibility for the entire software development life cycle. To enable and support these organizational changes, large and monolithic IT systems are being broken down into smaller applications. By reducing dependencies between applications, teams can work more independently, thus increasing the flexibility and speed of development.

However, the transition towards continuous development and smaller applications is challenging NAV's use of analytical data. Within NAV, analytical data has traditionally been managed by a single unit, the Knowledge department. As the name implies, the Knowledge department has been responsible for producing analytical insight about NAV, ranging from public statistics to internal steering information. By collecting data from various data sources and synthesizing them into a coherent model (data warehouse), the Knowledge department has been able to provide insight across business domains. But the centralized does not scale: As the number of data sources and change rates increase, the Knowledge department has become a bottleneck and a potential source of error. To manage these shortcomings, the NAV IT department has proposed a decentralized data management strategy, where teams take responsibility for preparing and sharing data produced by their applications.

In the following sections, we begin by giving a more detailed description of NAV's centralized data management strategy, and why it is incompatible with agile software development practice. We then continue to describe the ongoing transition towards decentralized data management and the challenges this entails.

### 4.2 Centralized data management

NAV is responsible for presenting statistics and steering information on welfare services and users. Among their customers are the Government, Statistics Norway, as well as the media and the public. Many of these statistics are regulated by law, including the Statistics act[3] and financial regulations[4]. The reported statistics are used for planning and prioritizing and influence internal operations as well as national interests.

*"NAV is a large enterprise, and it affects the stock market if our reports are wrong. What is happening [with the data] under our wings is of great importance nationally."*
(Member of Application development team).

---

[3] https://lovdata.no/dokument/NLO/lov/1989-06-16-54
[4] https://www.regjeringen.no/globalassets/upload/fin/vedlegg/okstyring/reglement_for_okonomistyring_i_staten.pdf

The Knowledge department has traditionally been responsible for gathering analytical data across NAV. These data have been extracted from source systems, transformed, and loaded into a data warehouse. The data warehouse team has been responsible for transforming and compiling data into a coherent data model. This requires extensive knowledge of both source systems and business domains:

*"[Data] must be arranged such that you don´t put apples and grapes in the same report. You need to understand the concepts which were in the data when they were originally reported. [...] This is addressed in the traditional data warehouse model, with ETL [Extract-Transform-Load] thinking and processes for extracting and transforming data, where you know with certainty what has happened to the data which lay in your centralized data storage."* (Member of data warehouse team).

This approach worked reasonably well when the system landscape consisted of large monolithic IT-systems and databases that rarely changed. As formulated by a member of the Data platform team:

*"Back then [two to three years back], the data warehouse team could extract all the data, and changes were quite rare. Because changes were a hassle".*

However, with the transition towards agile development and micro architectures, applications and databases began to change more frequently. For some systems, change rates increased from yearly to hourly releases.

The data warehouse team was unable to cope with the escalating number of changes, forcing the NAV to look for alternative data management strategies.

*"The centralized data warehouse environment cannot keep up with the pace because they are not rigged for it. It was doomed to fail before they tried, because somehow you suddenly have 150 applications instead of a few large monoliths. [...] We have gone from making changes [to our software] four times a year [...] to around 1300 times a week. In other words, continuous deployment, and it is no longer possible for a centralized environment to keep up with all the changes. Things break in pipelines and then things stop working and are not updated. So, this has been the big question: What do we do to fix it? How do we equip ourselves?"* (Product owner).

To address the problem, the IT department proposed a distributed model, described as a "data-mesh" [7], where application development teams take responsibility for creating products and sharing data.

### 4.3 Towards agile data management

The distributed data management model, or "data-mesh" [7], can be described in terms of 1) data products and ownership, 2) data platform, and 3) federated governance. Each of these elements, and their interpretation within NAV are described below.

**Data products and decentralized data ownership.** Foundational to the data mesh is the decentralization of data ownership. For NAV, a shift from centralized to decentralized ownership implies that application development teams assume responsibility for their own data:

*"It is not a technological change or a technical implementation that is the big change. The big changes come when we say to the teams, for example, the team working with unemployment benefits, that they are also responsible for producing analytical insight into the domain. Reporting and statistics. They don´t do this now, because today this is the responsibility of the Knowledge department"* (Member of the Platform team).

With the distributed data ownership, interpretations and decisions relating to the data are done by the people closest to the data. In addition to sharing data with other teams, the distributed ownership model is thought to increase the quality of analytical data within the team:

*"We not only want the teams to share [their data] with others. We also want the teams to become aware of the possibility of using these data themselves to make decisions. This will result in better data for everyone"* (Member of Data platform team).

As a means of implementing data ownership, teams will develop so called "data products". A data product is defined as a dataset and the documentation it. Data products require deliberate design and management, satisfying the needs of prospective users.

The term "data product" is used to show that data needs to be treated as other products or services within the organization, and that the team. This requires insight into the needs of prospective users, and a strategy for maintaining and improving the products.

However, the transition towards distributed data management causes concerns in some parts of the organization. One informant addresses the fear of losing control over the data:

*"We are concerned that when individual teams take ownership of data and begin to produce data products, we might lose oversight over the different domains. This means that it must be clear who has responsibility for what, which isn´t currently the case"* (Product owner, Data warehouse team).

Others were concerned that the teams neither has the competence nor the time to take responsibility for the data and that data consumers would no longer have insight into and control over the extraction and transformation.

*"Data won´t be prioritized. That's our experience. Developing data products is not something development teams usually think about when they develop systems. They are concerned with the [end] user, and how the case worker will use the system. Data is way down on their list"* (Product owner, Data warehouse team).

**Self-serve data platform.** To enable distributed data ownership, the organization has introduced a self-serve data platform called NADA. The new data platform differs from the data warehouse in several ways. Most importantly in the way data is shared: While data in the data warehouse is collected and curated by a single team, the new data platform offers functionality which allows all teams to share their data. The NADA platform is thus a multisided platform where the entire organization can produce and consume data.

Despite the need for alternative ways of managing analytical data, there is not yet consensus across the organization concerning the new data strategy. For distributed data management to be introduced, the IT department must therefore develop a data platform which simplifies data sharing and analysis, as compared to existing solutions:

*"If a team is to become responsible for publishing insights concerning their domain, then they must have tools that make it easy. How can they publish a data product that provides insight into changes [within the domain] over time, or the number of cases we have processed per day? How can you publish this information easily?"* (Member of platform development team).

The platform will become a marketplace where producers and consumers meet to exchange data. To increase the value of the platform, the platform development team actively encourages data producers to offer their data on the platform. The platform team describes this process of identifying needs and encouraging teams to add data products as "growth hacking" The platform team tries to understand the needs of users, and subsequently going out into the organization to get these needs fulfilled:

*"I ask teams that have data which I know will be useful to others to create data products and deploy them on the platform"* (Member of data platform team).

In addition to facilitating the creation of data products, the platform will have a dashboard and tools for analysis. The output of the analysis can in turn be used to create new data products, thus allowing insights to be shared and reused across the organization. The platform is based on Google Development Platform and data products are created in BigQuery[5]. Although BigQuery is currently the only available technology on the platform, the platform team plans to offer other technologies in the future.

However, developing a multisided platform is challenging, since there is no direct interaction between producers and consumers of data, and a producer is not directly rewarded for preparing and sharing their data.

*"With the data platform, on the other hand, you have two types of users: You have those who produce data and those who need data. We therefore use the metaphor ´data marketplace´. We are creating a marketplace where it should be possible to offer and to find data. So, it is a more complex image for us who create the platform because we are not simply a service provider. […] So, we have more of a chicken and egg problem, where you need some users on the consumer side, since this gives value. But to get some consumers, you also need some data which they can consume"* (Member of platform development team).

**Federated computational governance.** At the time of writing, very few rules govern the creation and dissemination of data products on the platform. This follows from the platform team's deliberate intension of minimizing the number of rules enforced:

*"As the data platform provider, we do not wish to become a large, centralized decision-maker. We wish to listen to our users to understand their needs, and we aim to be very restrictive with implementing rules"* (member of data platform team).

The creation of rules thus happens through ongoing negations, where rules are formed in collaboration with data producers and consumers:

*"So far, we don´t have many rules that apply, because we have very few users both on the consumer and producer side, but this is an ongoing discussion. How do we agree on the rules? For example, should we use one type of key to identify a person? Should we use birth number? Should that be the key for all data, or should each individual*

---

[5] https://cloud.google.com/bigquery

*domain be able to have its own? We have several keys identifying a person today. […]. These are rules we must agree on. But to know what [rules] to make, we need to know what users need. For this, I need a forum where producers and consumers of data can meet and agree on the rules"* (Member of platform team).

The IT department is exploring how they can maintain the privacy and security of citizens, while simultaneously stimulating teams to share and use data. To address this challenge, domain teams have access to a "privacy coach", which gives them legal counseling in the use of data. The IT department also has a "Data treatment catalog", where the use of sensitive data is recorded and justified. However, the data treatment catalogue has not yet been linked to the data platform:

*"All teams that treat data should register this treatment in the Data treatment catalogue and make the information available to the rest of NAV and to the authorities. It can also be used for other purposes, but so far, it is not linked to the [new] data platform. So, the ability to describe datasets and the legal justification for use has not yet been linked to the platform"* (Data analyst).

Whereas some data products only involved data from a single domain, the most valuable data products are those that involve multiple products and domains. One example of such domain-spanning products are unemployment figures. Unemployment figures cannot be calculated from a single system but are based on "all the things which a person is not". For instance, an unemployed person is not under education, is not on sick leave, and is not temporarily laid off, elements of information which is gathered from a series of different information systems. Other examples of compilations of data from multiple domains are average case processing time, and the number of erroneous payments made by NAV. Using data from different domains require knowledge of these domains. In the centralized model, this competence is held by the Knowledge department, and there is concern that cross-organizational insight and the ability to analyze data across domains will be reduced with distributed data management and local ownership of data.

*"It requires a lot of competence to use data from other domains. So, if you are to use data from another domain, it must be well documented. The data must be processed in a way that makes it easy to understand and user-friendly. In addition, what does it mean to connect data [from different domains]? This is a type of competence which takes time, and which must be acquired by the teams that work with source systems"* (Employee in the Knowledge department)

Another concern relates to the willingness of teams to invest in data products, as they have no direct benefit in sharing the data. Some informants therefore believe that data sharing must be compulsory:

*"We want there to be established requirements, compelling teams to make data available. And make sure that this data is made available as part of the statistics and steering information. Otherwise, it will be difficult for us, because we cannot involve ourselves with all the 120 teams"* (Product owner, Data warehouse team).

## 5      Discussion

Our initial involvement with NAV has provided some early insights regarding both the need and motivation for considering data mesh as a strategy for becoming data-driven, but also insights into challenges that follow from such a transition. NAV is the largest service organization in the country and administers data on – literally – every Norwegian citizen. However, technical, and organizational legacy is challenging the organization´s agility, and their ability to convert data into actionable insights.

NAV's journey towards increased agility has so far taken the organization through two transitions. First, the IT department enabled autonomous and cross-functional teams that build domain knowledge in product areas such as Work, Health, and Family. Cross-functional teams within each domain have autonomy and responsibility for the continuous development and deployment of related IT services, e.g., caseworker support systems.

Second, to match this way of organizing software teams, the system architecture has been transformed over time: Large and monolithic systems have been broken down into micro-services, enabling independent and loosely coupled applications that can be managed by single teams. However, although these transitions have increased the agility of the software development organization, they have also triggered the need for alternative ways of managing data. Traditional data management models, where analytical data are gathered in data silos and interpreted by a centralized unit, are unsuited for a distributed and continuous reality of agile software development. As monolithic systems are broken down into smaller applications, change rates have increase, and data pipelines are broken. Hence, applying the ideas of data mesh and distributed data management might be considered an imperative in further increasing the agility of software organizations and creating a data-driven organization.

The transition towards the principles of data mesh and distributed data management is however in an early stage and does not come without challenges. Our findings reveal that such a radical change creates uncertainties in various parts of the organization, depending on their need for and use of data.

**Challenge 1 – Change of control of data extraction and transformation.** Analysts in the knowledge department are concerned that they might lose control of data sources, resulting in erroneous data and reduced quality. They argue that having overview of the various domains is necessary when producing national statistics and providing analysis to national authorities and policy makers. They argue that such an overview cannot be obtained in application development teams, which role is precisely to specialize in a specific domain. The question thus remains: When the traditional centralized data management model where a single unit is responsible for gaining cross-organizational insight is replaced by local ownership and data products – how will the organization be able to support compiled data products which require cross-organizational insights? How should new needs for data be communicated to many data-controlling domain teams?

**Challenge 2 – Managing rightful and legal access to, and use of data.** The principle that domain teams are responsible for offering "their" data as data products via a data platform that everyone can access raises concerns regarding control and rightful use of data according to regulations on data protection. The General Data Protection Regulation (GDPR), which all the countries in the European Union (EU) and European Economic Area (EEA) are covered by, is a good example of such a regulation that may problematize the data mesh mentality. How does one incorporate the data minimization principle in GDPR, which says that you should only collect and process the minimum amount of data possible to fulfill your purpose, in the data mesh where one wishes to collect and process as much data as possible? Or how can one provide a transparent description of how a person´s data is processed as demanded by the lawfulness, fairness, and transparency principle in GDPR when the aim of the data mesh is to provide the data to everyone with the goal of continuously discovering new innovative ways of utilizing the data? The principles of data mesh and GDPR does not fully harmonize, and it is unknown how a data mesh should be managed in practice to reap the fruits of Dehghani´s [7] data mesh principles as well as being in line with the General Data Protection Regulation.

**Challenge 3 – Creating data products.** Traditionally, software product teams have given little thought to the data stored in their databases beyond their own use in the specific application that is developed. It has been the responsibility of data engineers and analysts in the central unit to gather and prepare data for analytical purposes – using the data warehouse as the main data storage. Developers have traditionally been focused on developing of end-user functionality, lacking both the competence and motivation for prepare and enrich data for analytical use. Hence, there is a need to add competence and capacity to the domain teams. It is however unclear which skillset it requires, and what the cost will be.

Related to this issue is the need for data products that span multiple product domains. Although some insight can be gained by analysing data within a single domain, the most valuable use cases involve a combination of data products from different domains. The autonomy of domains must therefore be combined with some degree of standardization, making it possible for data products to be combined. This requires insight into other business domains, as well as one's own.

Several questions therefore need to be answered: Should cross-functional teams include a data scientist function or role? How much can be automated and supported by the data platform? And how should a team learn about the need for data in other business domains?

**Challenge 4 - Establishing a thriving ecosystem.** A functioning data mesh builds on data owners that publish data products that can be consumed by others. But which incentives does a software team have to invest time in preparing, publishing, and maintaining data products? Of course, in a system where most teams can make use of data from other teams, we could foresee a naturally functioning ecosystem – but do we un-derstand such mechanisms properly? Should the publication of data products be an organizational obligation or are there other mechanisms that could be put into play? For example, could we make use of the same incentives that drive open-source development of code, where opening your data means that other's provide valuable feedback and enriched data in return? Would opening of data mean that the product team as a data provider put extra effort in making data understandable and useful – in short, establish proper quality of the data product?

This overview of challenges is not exhaustive and by no means complete. It merely provides an initial understanding of the many challenges related to the effectuation of the data mesh strategy in a complex and data-intensive software organization as perceived by informants in NAV. Had we studied other organizations within other sectors or countries, these challenges might differ.

There are also some challenges not addressed in the current study: Among these are whether it is possible to host the data platform on a cloud service by a service provider which is located outside of EU/EEA in a lawful way. The Schrems II ruling by the Court of Justice of the European Union states that "companies must verify, on a case-by-case basis, whether the law in the recipient country ensures adequate protection, under EU law, for personal data transferred under SCCs and, where it doesn't, that companies must provide additional safeguards or suspend transfers". The US, where most of the largest cloud providers have their headquarters, is a country that often is not considered a country where personal data is adequately protected.

We have studied a single case organization and a recent phenomenon (the transition towards a data mesh) in an early phase, over a restricted period (approximately 6 months). This naturally restricts generalizability. However, the study provides valuable early insights into a very large and complex organization that seeks to implement increase the use of efficiency of analytical data by introducing distributed data management – a challenge shared by many data-rich organizations. Furthermore, there are yet few reports from practice on how a "data mesh" can be realized and the challenges which organizations might face. We hope that others can build on this in future work. We have aimed to ensure validity by following acknowledged guidance on case studies [6] We have gathered data from more than one source (triangulation); document analysis (e.g. strategy documents) and interviews (covering a wide span in the IT organization), and we have collected data within a real-life context (NAV)

## 6　Conclusion

Our findings suggest that the organization agrees on the need for alternative ways of managing analytical data. There are however varying views on how this should be done, and how distributed data management and data mesh will affect the creation and use of analytical data. Also, although the main concepts, as laid out by Dehghani [7], are understood and motivates the transition, it is too early to see how these will be implemented, and how they will affect roles and work processes.

The ongoing transition is driven "inside-out", meaning that a data platform team offers a technical solution - the data platform, and supports teams that chose to take the platform into use. Some challenges have been identified and need to be addressed, while others have yet to appear. We hope that the potential benefits of more agile data management inspire researcher to investigate these approaches in the years to come.

### 6.1 Future work

We will continue to follow NAV and their transition towards becoming a data-driven organization. In that, we will 1) address the challenges that was identified in this study (as well as new emerging challenges), 2) collect and analyze data to investigate whether the new approach – data mesh – provides the effects that initially motivated the investments, and 3) describe the details on *how* NAV, as a complex organization, implements these principles. This briefly described research agenda has a potential for extending the knowledge on agile data management, and on how organizations can make better use of analytical data in improved insight and services. Furthermore, observing the case over time will give a basis for developing theories concerning the adoption of distributed data management. Leaning on the proposed framework by Eisenhardt [25], we have initiated some of the recommended steps, such as *Getting Started*, *Selecting Cases*, and *Entering the Field*.

## 7 Acknowledgements

This work was conducted in the project Transformit supported by the Research Council of Norway through grant 321477. The project was led and sponsored by the company KnowIT AS. We also want to thank NAV for granting us access to the case.